\documentclass[11pt]{article}
\usepackage{moriond,epsfig}

\def\Journal#1#2#3#4{{#1} {\bf #2}, #3 (#4)}

\def\NPB{{\em Nucl. Phys.} B}

\def\PRL{\em Phys. Rev. Lett.}
\def\PRD{{\em Phys. Rev.} D}

\def\AP{\em Annalen der Physik}
\def\IJMPA{{\em Int. J. Mod. Phys.} A}

\def\MPLA{{\em Mod. Phys. Lett.} A}
\def\CQG{\em Class. Quantum Grav.}
\def\AnJ{\em Astron. J.}
\def\AJ{\em Astrophys. J.}

\def\N{\em Nature}
\def\AA{\em Astron. Astrophys.}

\def\AIHPA{{\em Ann. Inst. H. Poincar\'e} A}
\def\ASR{\em Adv. Space Res.}

\def\SPAWB{\em Sitz. Preuss. Akad. Wiss. Berlin}
\def\NGWG{\em Nach. Ges. Wiss. G\"ottingen}
\def\MNRAS{\em Mon. Not. R. Astron. Soc.} 
\def\PIAUS{\em Proc. IAU Symposium}
\def\LRR{\em Living Rev. Rel.}
\def\PNP{\em Part. Nucl. Phys.}
\def\CP{\em Contemp. Phys.}
\def\EA{\em Exp. Astron.}

\def\eprint#1{{\it Preprint} #1}
\def\etal{\textit{et al }}

\def\url#1{{\rm #1}}

\def\be{\begin{equation}}
\def\ee{\end{equation}}
\def\bea{\begin{eqnarray}}
\def\eea{\end{eqnarray}}

\def\g{g}
\def\munu{{\mu\nu}}
\def\R{R}                  
\def\E{E}

\def\Pst{\phi}

\def\t{t}                  
\def\s{s}                  
\def\xb{\mathbf{x}}       
\def\xbx{\xb_1}
\def\xby{{\xb_2}}
\def\r{r}                  
\def\rx{{\r_1}}
\def\ry{{\r_2}}
\def\rydd{{\ddot \r}_2}
\def\ri{\rho}              
\def\th{\theta}            
\def\ang{\varphi}
\def\angxy{\phi}

\def\k{k}                  
\def\ad{a}                 

\def\T{T}
\def\pro{\pi}              

\def\td{{\cal T}}

\def\sec{{\rm sec}}
\def\ann{{\rm ann}}

\def\c{c}                  
\def\M{M}                  
\def\G{G}
\def\aP{a_{P}}

\def\stand#1{\left[#1\right]_\mathrm{GR}}
\def\unit#1{{\rm #1}}

\begin{document}
\vspace*{4cm}
\title{TESTS OF GRAVITY AT THE SOLAR SYSTEM SCALE}

\author{M.-T. JAEKEL}
\address{Laboratoire de Physique Th\'eorique de l'Ecole Normale Sup\'{e}rieure, 
CNRS, UPMC, 24 rue Lhomond, F75231 Paris Cedex 05}

\author{S. REYNAUD}
\address{ Laboratoire Kastler Brossel, Universit\'{e} Pierre et
Marie Curie, case 74, CNRS, ENS, Campus Jussieu, F75252 Paris Cedex 05}

\maketitle
\abstracts{As confirmed by tests performed  in the solar system,  General Relativity (GR)
presently represents the best description of gravitation. 
It is however challenged by observations at very large length scales, and already at the solar
system scale, tracking of the Pioneer 10/11 probes  has failed 
to confirm their expected behavior according to GR. Metric extensions of GR, which are presented here, have the quality
of preserving the fundamental properties of GR while introducing scale dependent modifications. 
We show that they moreover represent an appropriate family of gravitation theories to be compared with observations
when analysing gravity tests.   
 We also discuss different tests which  
could allow one to determine the metric extension of GR prevailing  in the solar system.
}

\section{Introduction}
 General Relativity (GR) is unique among fundamental theories as 
it has first been introduced on the basis
of general principles \cite{Einstein16}, before being confirmed by 
observations \cite{Will05}. However, while GR agrees with the most precise observations made in the solar system,  recent observations performed at larger length scales show inconsistencies between
the visible content of larger parts of the Universe and the gravitation laws according to GR. 
The anomalous rotation curves of galaxies \cite{McGaugh11} and the anomalous acceleration of type Ia supernovae \cite{Perlmutter99} can point at the existence of important amounts of dark matter
in galactic halos \cite{Gentile04,Bienayme11} and of dark matter and energy at the cosmological scale \cite{Ferramacho09,Blanchard11}.
But, should these dark constituents remain unobserved, 
this could mean that the gravitation laws have to be changed at these scales.
The necessity to modify GR may even come earlier, already at the solar system scale, if the anomaly 
observed on the navigation data of the Pioneer 10/11 probes \cite{Anderson98}
did not find a conventional explanation.     
 
Beside observational data, theoretical arguments also plead for considering the possibility of 
scale dependent gravitation laws. 
The coupling constants of the other three fundamental interactions  
are known to develop a scale dependence as a consequence of radiative corrections, a property which justifies the idea of a possible unification of all fundamental interactions. Gravitation, being also  both geometry and a field theory, should share this property. Assuming
"asymptotic safety"  \cite{Weinberg79}, renormalization group techniques allow one to derive  
the general features of the scale dependence of gravitation.
When combined with observational constraints, they lead to favour a family of metric extensions 
of GR for describing gravitation \cite{JR07}.
  
We briefly review here the properties of such metric extensions of GR
and obtain a parametrization of these theories suiting phenomenological purposes.
We discuss how they can be used when analysing gravity tests performed 
in the solar system and when searching anomalous gravitation properties with respect to GR.

\section{General Relativity and its metric extensions}
GR plays an exemplary role among fundamental theories because of two essential properties:
it describes gravitation both as geometry and as a field theory. 
The first property deeply affects modern spacetime metrology, which 
relies on a strong relation between gravitation and  geometry: definitions of reference systems depend on an underlying metric $\g_{\mu\nu}$ which refers to solutions of gravitational equations of motion \cite{IAU2000}. 
This assumption is made possible by the identification of gravitation with the geometry of spacetime. 
According to GR, all bodies, massive and massless ones as well, follow geodesics in absence of 
non gravitational forces. 
Geodesics are obtained from a universal geometric distance, defined by the metric $\g_{\mu\nu}$ and
which also coincides with the proper time delivered by clocks along their motions. 
This results in particular in the universality of free fall, a principle which has been verified to hold 
at very different length scales, ranging from millimeter \cite{Adelberger09,Adelberger11} to astronomic scales \cite{Williams04}, and  at a very high precision level ($10^{-13}$).
 
On the other hand, as one of the four fundamental interactions, 
gravitation is also described by means of a field, 
 characterized by the way it couples to its sources.
In GR, the metric field couples to energy-momentum tensors $\T_\munu$ through its Einstein curvature $\E_{\munu}$,
a particular combination of Ricci ($\R_\munu$) and scalar ($\R$) curvatures. 
As both tensors are divergenceless,  $\T_\munu$ as a consequence of
 conservation laws and $\E_{\munu}$ of Bianchi identities, coupling can be realized by a unique
proportionality constant, Newton gravitation constant $G_N$ \cite{Einstein15,Hilbert15} ($c$ is light velocity)
\bea
\label{GR_gravitation_law}
\E_\munu \equiv \R_\munu-{1\over2}\g_\munu \R={8\pi\G_N \over\c^4}\T_\munu
\eea
But gravitation is a very weak interaction, so that the particular form of
the gravitational equations of motion (\ref{GR_gravitation_law}) is extremely difficult
to bring to  experimental test. Usually, tests of gravity are only performed in an indirect way,
by comparing observations with predictions which can be obtained on the basis of metrics satisfying 
equations (\ref{GR_gravitation_law}). 
As a consequence, the particular field theory characterizing gravitation, and $G_N$,
 appear to be tested with much less precision than the geometric nature of gravitation.
   
Moreover, theoretical arguments suggest that the gravitational equations of motion 
(\ref{GR_gravitation_law}) cannot remain valid
over arbitrary energy or length  scales. Indeed, as a universal mechanism occuring in field theories, 
higher order processes modify couplings and propagators. 
This is the case for electro-weak and strong interactions, whose coupling constants become scale dependent and 
follow renormalization group trajectories. 
In a similar way, radiative corrections should lead to a scale dependence of the gravitational
coupling, making the gravitational equations specified by GR (\ref{GR_gravitation_law})
only approximately valid \cite{Weinberg79}.  Remarkably, these theoretical arguments appear to be  met by anomalous observations performed at very large length scales \cite{McGaugh11,Perlmutter99}, which can also be interpreted as questioning the validity of GR at such scales \cite{Gentile04,Bienayme11,Ferramacho09,Blanchard11}. 
 
Although the case of gravitation shows to be theoretically involved,
 the main features of the expected  scale dependences can nonetheless be
obtained from general properties. 
The symmetries, or gauge invariance, underlying gravitation constrain observables to 
take the form of geometric quantities \cite{Sakharov67,tHooft74}. 
Hence, the further couplings induced by radiative corrections
involve squares of curvatures so that GR can indeed  be seen to  be embedded in a family
of renormalizable field theories.   
This implies that, when radiative corrections are taken into account, gravitation can still be described by a metric theory, but that the single gravitation constant $G_N$ must be replaced by several running coupling constants \cite{Fradkin82} characterizing
additional terms in the Lagrangian. 
 There results that GR, defined 
by Einstein-Hilbert Lagrangian  (\ref{GR_gravitation_law}), is extended to a theory which 
is both non local, as a result of radiative corrections,
and non linear, due to their geometric nature.
It leads to gravitational equations of motion which can be put under a general form, with  a 
 susceptibility  replacing Newton gravitation constant \cite{JR95} 
\bea
\label{general_gravitation_law}
\E_\munu= \chi_\munu (\T)=  \frac{8\pi G_N}{c^4} T_\munu + \delta \chi_\munu(T)
\eea
The resulting equations appear to be difficult to solve due to a particular mixing realized
 between non linearity and non locality.  

As another general property, radiative corrections can be seen to
essentially differ in two sectors corresponding to couplings to massless
or massive fields \cite{Fradkin82,JR95}:
in the former case, traceless energy-momentum tensors couple to Weyl curvature only,
while in the latter case couplings between energy-momentum traces and the scalar curvature also occur.
GR should then be extended to metric theories which are characterized by two sectors,
of different conformal weights,
with corresponding running coupling constants $G^{(0)}$ and $G^{(1)}$ which
generalize Newton gravitation constant $G_N$. 
The relations between coupling constants can be given simple expressions in a linearized approximation
(using a representation of fields in terms of momentum $k$ and introducing
the corresponding projectors $\pro$ on trace and traceless parts)\cite{JR05mpl,JR05cqg}
\bea
\label{linear_decomposition}
&&\E_{\mu\nu}= \E^{(0)}_{\mu\nu} + \E^{(1)}_{\mu\nu}, \qquad \pro_{\mu\nu}\equiv \eta_{\mu\nu} -{\k_\mu \k_\nu\over k^2}\nonumber\\
&&\E^{(0)}_{\mu\nu} = 
\lbrace \pro _{\mu}^{0}\pro_{\nu}^{0}-{\pro_{\mu\nu}\pro^{00}\over3}\rbrace
\,
\frac{8\pi \G^{(0)}}{\c^{4}}\T_{00}, \qquad 
\E^{(1)}_{\mu\nu} =  \frac{\pro _{\mu\nu}\pro ^{00}}{3}
\frac{8\pi \G^{(1)}}{\c^{4}}\T_{00}  \nonumber\\
&&\G^{(0)}= \G_N  + \delta\G^{(0)}, \qquad
\G^{(1)}= \G_N  + \delta\G^{(1)}
\eea
Although the two running coupling constants remain close to $G_N$,
non locality and non linearity combine in an intricate way and
do not allow a decomposition as simple as (\ref{linear_decomposition}) to hold 
beyond the linearized approximation. Alternatively, one can look for
non linear but local theories which approximate the previous metric extensions of GR.
 It is remarkable that, due to the presence of two sectors, such approximations
can be obtained which involve higher order field derivatives and nonetheless correspond to theories
with stable ground states \cite{Bruneton07}.  
 
To the theoretical difficulties implied by non locality combined with non linearity,
some compensation can be found in direct observations. Indeed,    
the latter show that gravitation should remain very close to GR over a large range
of scales. They moreover show that departures from GR can happen not only at 
large energy scales,
as expected if gravitation should unify with other fundamental interactions, but also 
at large length scales \cite{McGaugh11,Perlmutter99}. 
Gravitation tests performed up to now  make it legitimate to consider the effective gravitation theory at ordinary macroscopic length scales to 
be a perturbation of GR \cite{Will05}. Solutions of the generalized  equations  (\ref{general_gravitation_law}) should then correspond to perturbations of the solutions of GR equations of motion (\ref{GR_gravitation_law}). Equivalently,
equations (\ref{general_gravitation_law}) may  be seen as providing 
metrics which remain close to those determined by GR and just differ from the latter by 
curvature anomalies \cite{JR06cqg}
\bea
\label{perturbed_GR}
&&\E = \stand{\E} + \delta \E, \qquad
\stand{\E} = 0 \qquad {\rm{where}} \qquad \T \equiv 0\nonumber\\ 
\label{Einstein_curvature}
&&\delta\E = \delta \E^{(0)}+\delta \E^{(1)}
\eea
Metric extensions of GR are thus characterized by two independent components 
of Einstein curvature tensor $\delta \E^{(0)}$ and $\delta\E^{(1)}$,
 reflecting the two different running coupling constants $G^{(0)}$ and $G^{(1)}$ modifying $G_N$ 
(as seen in the linear approximation  (\ref{linear_decomposition})).  When solving the gravitation equations of motion (\ref{general_gravitation_law}), the two independent Einstein curvature components are
replaced by two gauge-invariant potentials $\Phi_N$ and $\Phi_P$ 
(for a point-like source, using Schwarzschild coordinates) 
\bea
\label{metric_extension}
\delta E^0_0\equiv 2 u^4(\Phi_N -\delta\Phi_P)^{\prime\prime}, 
\quad \delta E^r_r \equiv 2 u^3\Phi_P^\prime \qquad u\equiv\frac{1}{r}, \quad
  ()^\prime\equiv \partial_u
\eea
In the case of GR, Einstein curvature vanishes and the solution depends on a single potential $\Phi_N$
taking  a Newtonian form  ($\Phi_P$ vanishing in this case).
In the general case, the potential $\Phi_N$ extends Newton potential 
while $\Phi_P$ describes a second gravitational sector. These potentials
can be seen as a parametrization of admissible metrics in the vicinity of GR solutions, 
which thus represent good candidates 
for extending GR beyong ordinary macroscopic scales. This parametrization appears to be  appropriate 
for analysing existing gravity tests and confronting GR with plausible alternative theories of gravitation.

\section{Phenomenology in the solar system and gravity tests}
The solution of the gravitation equations of motion  (\ref{general_gravitation_law}) takes a simple form
in the case of a stationary point-like gravitational source,
as it corresponds  to a static isotropic metric which reduces to two independent components
(written here in  spherical isotropic coordinates)
\bea
\label{isotropic_metric}
d\s^2 &=& \g_{00} \c^2 d\t^2 + \g_{\r\r} \left( d\r^2  +
\r^2(d\th^2 + {\rm \sin}^2\th  d\ang^2) \right)\nonumber\\
\g_{00} &=& \stand{\g_{00}} + \delta \g_{00}, \qquad 
\g_{rr} = \stand{\g_{\r\r}} + \delta \g_{\r\r}\
\eea
$\stand{\g}$ denotes the approximate metric satisfying GR equations of motion (\ref{GR_gravitation_law}),
which can be written in terms of Newton potential. In this case,  
the two independent components of the metric $\delta\g_{00}$ and $\delta\g_{\r\r}$
are in one to one correspondence with the two independent components of Eintein curvature
 $\delta \E^{(0)}$ and $\delta\E^{(1)}$. Quite generally, the explicit expressions of the metric components 
in terms of the two gravitational potentials $\Phi_N$ and $\Phi_P$ (\ref{metric_extension})
are obtained by inverting the usual relation between metrics and curvatures \cite{JR06cqg}.

The most precise tests of GR  have been realized in the solar system.
Phenomenology in the solar system is usually performed with parametrized
post-Newtonian (PPN) metrics \cite{WillNordtvedt72,NordvedtWill72}.
Neglecting the Sun's motions, the corresponding PPN metrics reduce to the
form (\ref{isotropic_metric}), with $\g_{00}$ and $\g_{\r\r}$ being determined by Newton potential
$\Pst$ and two Eddington parameters $\beta$ and $\gamma$
\bea
\label{GR_solar_metric}
\g_{00} &=& 1+2\Pst+2\beta\Pst^2+\ldots\quad, \quad
\g_{rr} = - 1+2\gamma\Pst+\ldots \nonumber\\
\Pst &\equiv& -{\G_N\M\over\c^2\r}, \qquad 
\left\vert\Pst\right\vert \ll 1
\eea
The parameters $\beta$ and $\gamma$ describe deviations from GR (obtained for $\beta=\gamma=1$)
in the two sectors, corresponding 
respectively  to effects on the motion of massive probes and on light deflection.
PPN metrics are a particular case of metric extensions of GR, 
corresponding to a two-dimensional family which describes non vanishing but short range Einstein curvatures 
\bea
\label{PPN_case}
&&\Phi_N = \Pst + (\beta-1)\Pst^2 + O(\Pst^3), \qquad 
\Phi_P = -(\gamma-1)\Pst + O(\Pst^2)\nonumber\\
&&\delta\E^0_0 = {1\over\r^2} O( \Pst^2),\qquad
\delta\E^\r_\r = {1\over\r^2} \left(2 (\gamma-1) \Pst +O( \Pst^2) \right)\qquad \mathrm{[PPN]}
\eea
In contrast, general metric extensions of GR are parametrized by two gravitational potentials $\Phi_N$ and $\Phi_P$ (\ref{metric_extension}) describing arbitrary Einstein curvatures. These two functions
may be seen as promoting the constant parameters $\beta$ and $\gamma$ to  scale dependent functions. 
The latter manifest themselves as an additional dependence of gravitational effects on a geometric distance.
 The latter can be either a distance between points (as the probe and the
gravitational source) or a distance between 
a point and a geodesic (as the impact parameter of a light ray).    

Existing gravity tests put constraints on possible deviations from GR, hence on
allowed metric extensions of GR (\ref{isotropic_metric}) at the scale of the solar system.
Direct scale dependence tests have up to now been performed in the first sector only.
They were designed to look for possible
modifications of Newton potential taking the form of a Yukawa
potential ($\delta \Pst(\r) = \alpha e^{-\frac\r\lambda} \Pst(\r)$), characterized by a stength parameter $\alpha$ and a range $\lambda$.
These tests, performed for $\lambda$ ranging from the submillimeter range,
using dedicated experiments \cite{Fischbach98,Reynaud11},
to the range of planetary orbits, using probe navigation data and planetray ephemerides \cite{Fischbach98,RJ05}, 
show that the strength $\alpha$ of a Yukawa-like perturbation must remain rather small at all these scales,
so that the form of the gravitational potential in the first sector is rather strongly constrained to remain Newtonian. However, constraints become much less stringent below the submillimeter range, where 
Casimir forces become important \cite{Reynaud11}, and at scales of the order of the outer solar system, where 
 observations used to determine ephemerides become less precise. They moreover only concern the first sector.

The increasing set of observations performed in the solar system has progressively
 reduced the allowed deviations from GR
for the two PPN parameters $\beta$ and $\gamma$.   
Presently, the best constraint on the value of $\gamma$ is given by the measurement of the Shapiro
time delay, induced by the gravitational field of the Sun on the radio link which was used to follow
 the Cassini probe during its travel to Saturn \cite{Bertotti03}.
GR prediction for the variation of the deflection angle, near occultation by the Sun, has been
confirmed, constraining  $\gamma$ to be close to $1$  with a precision of $2.5\times 10^{-5}$.
A similar bound is provided by VLBI measurements of light deflection \cite{Will05}.  
One may remark that such a precision is obtained when assuming that the parameter $\gamma$ remains
 constant. As the deflection angle decreases with the impact parameter of the ray,
the precision on the measurement of the deflection angle is mainly due to small impact parameters.
As a result, the corresponding constraints should be sensitively less stringent when confronted
 to general metric extensions of GR, which allow $\gamma$ to depend on the impact parameter of the ray.    

The value of $\gamma$ being assumed, the parameter $\beta$ can be obained either by means of a direct measurement, such as Lunar Laser Ranging \cite{Williams04}, measuring the Sun polarization effect on the 
Moon orbit around the Earth, or by means of big fits, using all data made available by probe navigation and astrometry measurements, to determine planet ephemerides
\cite{Folkner10,Pitjeva10,Fienga11}. Both methods lead to similar constraints on $\beta$, fixing the latter to remain close to $1$, up to deviations less than $10^{-4}$.   
Let us remark that these determinations are performed at the scale of the Moon orbit in one case, and at
a scale of several astronomic units ($\unit{AU}$) in the other case. 
They can also be considered as independent estimations
of $\beta$ being performed at different length scales.

Available data for gravitation at large length scales in the solar system are rather few. 
Hence, the navigation data of the Pioneer 10/11 probes, during their travel in the outer part of the solar system, provide an important consistency check for  models of gravitation in the solar system.
 Remarkably, the analysis of Doppler data has failed to confirm the predictions made according to GR. 
Comparison of observed with predicted values resulted in residuals which did not vanish but 
could be interpreted as exhibiting the presence of an anomalous acceleration $\aP = (0.87 \pm 0.13) ~\unit{nm}~\unit{s}^{-2}$, directed towards the Sun or the Earth, and approximately constant over distances ranging from $ 20\unit{AU}$ to  $70\unit{AU}$ \cite{Anderson98}. 
Many attempts have been made to find a conventional explanation to the Pioneer anomaly as a systematic effect
either related to the probe itself, allowed by a loss of energy from power generators on board,
or to the environment of the probe, due to the presence of dust or gravitating matter in the outer solar system
\cite{Nieto07}. These have been followed by sustained efforts for recovering further data
and performing new analyses covering the whole Pioner 10/11 missions \cite{Turyshev10}. Up to now, 
these attempts have remained unsuccessful in explaining the totality of the Pioneer anomaly. 

Furthermore, a recent study, confirming the secular part of the Pioneer anomaly, has also  
analysed the modulations apparent in the Doppler data, showing 
 that their frequencies correspond to the Earth's motions,
and that the Doppler residuals can be further reduced by introducing 
simple modulations of the radio links \cite{Levy09}.
Modulated anomalies cannot be produced by a conventional explanation of the secular part
but require a further mechanism (trajectory mismodeling, solar plasma effects, ...) 
to be accounted for. On the other hand, simple models modifying the metric are able to reproduce both types of anomalies. These features leave the  possibility of a common gravitational origin of the Pioneer anomalies,
pointing at a deficiency of GR occuring at length scales of the order of the solar system size.

\section{Tests of metric extensions of GR}
 Besides being favoured by theoretical arguments, metric extensions of GR also 
provide an appropriate tool for analysing  gravity tests performed in the solar system.
Most precise tests realized at or beyond the AU scale strongly rely on Doppler ranging, hence on an appropriate modeling of electromagnetic links and the trajectories of massive bodies. Metric extensions of GR provide a general and simple expression for the time delay function $\td(\xbx, \xby)$ which describes the links used to follow a massive probe
 ($\xb_a \equiv \r_a (\sin\th_a\cos\ang_a,\sin\th_a\sin\ang_a,\cos\th_a)$ with $a=2,1$ respectively 
denoting the coordinates of the probe and a station on Earth, $\td(\xbx, \xby)$ is written here for a static istropic metric (\ref{isotropic_metric})) \cite{JR06CQG}
\bea
\label{time_delay}
&&\c\td(\rx, \ry, \angxy) \equiv \int_\rx^\ry {-{\g_{rr}\over \g_{00}}(\r)
d\r \over \sqrt{-{ \g_{rr}\over \g_{00}}(\r) - {\ri^2\over\r^2}}}
\quad,\quad
\angxy = \int_\rx^{\ry} {\ri d \r/\r^2 \over
\sqrt{-{\g_{rr}\over\g_{00}}(\r) - {\ri^2\over\r^2}}}\nonumber\\
&&\cos\angxy\equiv\cos\th_1\cos\th_2+\sin\th_1\sin\th_2\cos\left(\ang_2-\ang_1\right) 
\eea
 $\angxy$ is the relative angle, when seen from the gravitational source,
 of the two points $\xb_1$ and $\xb_2$ and $\ri$ the impact parameter of the light ray
joining these points.
The two-point function $\td(\xbx, \xby)$ describes the time taken by a light-like signal to propagate from position $\xb_1$ to position $\xb_2$ (thus giving a parametrization of lightcones).
The time delay function can be seen to be parametrized  by  metric components (\ref{time_delay}), 
hence by the two gravitation potentials $(\Phi_N, \Phi_P)$. Doppler signals are obtained by taking the time derivative of $\td(\xbx, \xby)$, and evaluating the latter on the trajectories of the probe and 
the Earth station.
As geodesics must be determined according to the same metric extension of GR, 
 the two potentials also enter the expressions of the trajectories \cite{JR06cqg}.
  
Comparison between metric extensions and  GR predictions can be performed explicitly and analysing the former within the framework of GR  leads to deviations which take the form of 
Pioneer-like anomalies ($\delta\ad = \delta\ad_\sec + \delta\ad_\ann$ denotes the time derivative of Doppler signals \cite{Anderson98})
\bea
\label{simplified_Pioneer_anomaly}
&&\delta\ad _\sec \simeq -{\c^2\over2} \partial_\r(\delta\g_{00})
+\stand{\rydd}\left\lbrace{\delta(\g_{00}\g_{\r\r})\over2} -\delta\g_{00}\right\rbrace
-{\c^2\over2}\partial_\r^2\stand{\g_{00}}\delta\ry \nonumber\\
&&\delta\ad _\ann \simeq {d\over d\t}\left\lbrace\stand{{d\angxy\over d\t}}\delta\ri\right\rbrace 
\eea
The gravitational potentials in the two sectors contribute to both the secular part $\delta\ad_\sec$
and the modulated part $\delta\ad_\ann$ of the anomaly.
 These furthermore depend on the probe and Earth motions, which are obtained from the  
equations for geodesics and initial conditions. 
Hence, Pioneer-like anomalies appear as a prediction of metric extensions of GR.
These moreover predict strong correlations between secular and modulated anomalies, 
which can  be considered as signatures to be looked for in observations \cite{JR05cqg,JR06CQG}.

Besides directly, through a precise analysis of probe navigation data, the two gravitational
potentials may also be expected to be determined as part of a big fit of all navigation and astrometric data, such as those used to obtain the ephemerides of planets and some of their characteristic 
constants. In such an approach, the two potentials play the same role as
Eddington parameters $\beta$ and $\gamma$  \cite{Folkner10,Pitjeva10,Fienga11}, with the additional feature of allowing significant dependences on length scales of the order of the solar system
size \cite{JR06cqg}. The results of Doppler and ranging observations should then be taken into account
by using the time delay function (\ref{time_delay}) and the geodesics, depending on the two gravitational potentials $(\Phi_N, \Phi_P)$ which define a general metric extension of GR. 
Clearly, the need to recall to  numerical methods entails that the neighborhood defined by the two potentials, in their general form, is too large to be totally scanned by a fit.
Hence, it appears crucial to design simplified models which depend on a small number of real parameters
but still preserve the scale dependences which are most likely to be observed \cite{Hees11}.

Metric extensions of GR also predict effects which can be expected to be exhibited by future experiments benefiting from a  high increase in  precision measurement. 
The time delay function (\ref{time_delay}) results in a particular scale dependence of the  
gravitational deflection of light which can be equivalently represented as an additional dependence of
the deflection angle $\omega$, or else of Eddington parameter $\gamma$,  on the impact parameter 
of the light ray   ($M$ denotes the mass of the gravitational source, $r$ its distance to the observer,
$\chi$ the apparent relative angle between the light source and the gravitational source) 
\cite{JR05cqg}     
\bea
\omega(\chi)\simeq \frac{G_NM}{c^2 r}
\frac{1+\gamma(\chi)}{tan\frac{\chi}{2}} \nonumber
\eea
The two gravitational potentials characterizing metric extensions combine
to induce a modification of the deflection angle which, in contrast to GR,
 contains a part which increases with the impact parameter. Such deviations should then become more noticeable for
measurements performed with a high precision and at small deflection angles.
In a near future, GAIA will perform a survey of our neighborhood in our galaxy and will follow
 with a very good accuracy an extremely large number of astrometric objects \cite{GAIA01,Bienayme11}.
This will include in particular a very large number of light deflection observations 
performed at small deflection angles, or at large angular distances from the Sun
\bea
\delta \omega < 40 \mu{\rm{as}}, \qquad 
\omega  \sim 4 {\rm{mas}}, \qquad
\chi \in [45^\circ, 135^\circ]\nonumber
\eea
As a consequence, GAIA data will improve the accuracy for the observed mean value of $\gamma$ 
(better than $2\times 10^{-6}$) and will make it possibe to map the dependence of  $\gamma$ on $\chi$
over its whole range of variation. Such a mapping could put into evidence small deviations from GR and moreover allow to determine and fit their particular dependence.   

A definite answer to the question of modifying the gravitation theory at the solar system scale
would be provided by missions embarking dedicated means for directly measuring the effects of gravity.
 A first example is OSS mission \cite{OSS09} which, beside ranging facilities, will also possess a high precision accelerometer, thus allowing to distinguish the effects
of gravitation from other forces affecting the probe and hence to determine unambiguously
whether the probe follows a geodesic, and whether the latter corresponds to GR.
Another mission, SAGAS \cite{SAGAS}, aims at reaching the outer part of the solar system
with, beside an accelerometer, an atomic clock on board. 
Using the combined information obtained, with a very high precision, from the  optical links and
the clock on board, one would be able to reconstruct the gravitational potentials in the two sectors,
and thus to exactly determine the gravitation theory prevailing at the largest scales which can be reached by artificial probes.

\section{Conclusion}

When generalized under the form of a metric extension, GR remains a successful theory of gravitation
within the whole solar system. Minimal modifications  allow one to 
account for all gravity tests performed up to the solar system scale and to
confirm the position of GR as the basis of gravitation theory.
They moreover correspond to scale dependences of the gravitational coupling, 
thus bringing gravitation closer to the other fundamental interactions.

From a phenomenological point of view, metric extensions of GR appear as a convenient tool 
for testing gravity within the solar system. They may also provide
a natural answer to the presence of anomalies when  observations are analysed by confrontation
with GR.  The actual theory of gravitation can be approached by looking for such anomalies
occuring in residuals of direct ranging data or big fits. It may also be determined by future 
high precision observations (GAIA) or dedicated missions in the solar system (OSS, SAGAS).

\end{document}